# Weyl nodal line induced pairing in Ising superconductor and high critical field

Xiaoming Zhang[1,*] and Feng Liu[2,*]

[1] *Department of Physics, College of Information Science and Engineering, Ocean University of China, Qingdao, Shandong 266100, China*

[2] *Department of Materials Science and Engineering, University of Utah, Salt Lake City, Utah 84112, USA*

*Correspondence to: fliu@eng.utah.edu, zxm@ouc.edu.cn

**Superconducting and topological states are two quantum phenomena attracting much interest [1-3]. Their coexistence may lead to topological superconductivity sought-after for Majorana-based quantum computing [4-11]. However, there is no causal relationship between the two, since superconductivity is a many-body effect due to electron-electron interaction [12] while topology is a single-particle manifestation of electron band structure [13]. Here, we demonstrate a novel form of Fulde-Ferrell-Larkin-Ovchinnikov (FFLO) pairing [14,15], induced by topological Weyl nodal lines in Ising Bardeen-Cooper-Schrieffer (IBCS) superconductors. Based on first-principles calculations and analyses, we predict that the nonmagnetic metals of $MA_2Z_4$ family [16,17], including $α_1$-TaSi$_2$P$_4$, $α_1$-TaSi$_2$N$_4$, $α_1$-NbSi$_2$P$_4$, $α_2$-TaGe$_2$P$_4$, and $α_2$-NbGe$_2$P$_4$ monolayers, are all superconductors. While the intrinsic IBCS paring arises in these non-centrosymmetric systems [18-24], the extrinsic FFLO pairing is revealed to be evoked by the Weyl nodal lines under magnetic field, facilitating the formation of Cooper pairs with nonzero momentum in their vicinity. Moreover, we show that the IBCS pairing alone will enhance the in-plane critical field $B_c$ to ~10-50 times of Pauli paramagnetic limit $B_p$, and additional FFLO pairing can further triple the $B_c/B_p$ ratio. It therefore affords an effective approach to enhance the robustness of superconductivity. Also, the topology induced superconductivity renders naturally the possible existence of topological superconducting state.**

Superconductivity [1] and electron topology [2,3] are two landmark breakthroughs in the fields of condensed matter physics and material science. When they are brought together by proximity effect [4,5], or coexist in one material, e.g., superconductors with topological states [6-8] and *vice versa* [9], a more exotic quantum state of topological superconductivity arises, offering a promising route to fault-tolerant Majorana-based quantum computing [10,11]. However, conventional wisdom tells that there is no causal relationship between superconductivity and electron topology. This is understandable because the former is a many-body effect manifesting an attractive electron-electron interaction of Cooper pairs [12], while the latter is a single-particle effect induced by parity inversion in electron band structure [13]. Interestingly, we reveal a novel form of Fulde-Ferrell-Larkin-Ovchinnikov (FFLO) superconducting pairing [14,15], induced by topological Weyl nodal lines in the family of two-dimensional (2D) Ising superconductors of $MA_2Z_4$ monolayers [16,17]. It not only sheds new light on our fundamental understanding of superconductivity in relation with topology, but also provides a promising approach to enhance the robustness of superconductors.

In addition to critical transition temperature ($T_c$), another important figure of merit for superconductivity is critical magnetic field ($B_c$) beyond which the superconductivity vanishes. Generally, magnetic field destroys superconductivity through orbital and/or Pauli paramagnetic mechanisms. Because the orbital effect is weak or absent in those materials with large electron mass [25] or low dimensionality [26], suppressing Pauli effect has been the focus to increase $B_c$. In particular, the FFLO pairing [14,15,27-32] has been long shown as a feasible mechanism to enhance $B_c$ beyond the Pauli paramagnetic limit $B_p$. The formation of FFLO pairs, with non-zero momentum, stems from the spin-non-degenerate Fermi surfaces (FSs) induced by external magnetic field [27,28]. They are favored by low-dimensionality and anisotropic FS, tending to infinity at 1D limit at low temperature. The FFLO pairing has been mainly found in quasi-2D clean-limit superconductors, such as organic [28], Cuprate [29], iron-based [30], heavy-fermion superconductors [27], and van der Waals (vdW) layered $NbS_2$ [31,32].

On the other hand, recent studies have shown significantly enhanced $B_c$ in 2D Ising Bardeen-Cooper-Schrieffer (IBCS) superconductors [18-24,33-36], which suppresses the Pauli pair-breaking effect by an effective out-of-plane Zeeman field $B_{eff}$, induced by spin-orbit-coupling (SOC) together with broken inversion symmetry (Type-I) [18-21] or multiple degenerate orbitals (Type-II) [33-36]. Such mechanisms have been identified in transition-metal dichalcogenides (TMD) monolayers [18-24,34], few-layer stanene [35] and Pb films [36], to increase $B_c$ several times over $B_p$. Here, we demonstrate the coexistence of both FFLO and IBCS paring in the Type-I Ising superconductor of 2D monolayers with the formula of $MA_2Z_4$. Most interestingly, the 1D FFLO pairing, as triggered and confined by Weyl nodal lines, has the maximum stability [27,28].

It is worth mentioning that centimeter-scale monolayer films of $MoSi_2N_4$ and $WSi_2N_4$ have already been synthesized recently by chemical vapor deposition [16], which opens up a large family of 2D vdW layered materials with the general formula of $MA_2Z_4$ [16,17]. Extensive computational research shows that the $MA_2Z_4$ monolayers generally exhibit outstanding mechanical, thermal, electronic, optical, piezoelectric, thermoelectric, optoelectronics, and photocatalytic properties [16,17,37-42]. Of particular interest to us, certain $MA_2Z_4$ compounds have been theoretically predicted to be intrinsic superconductors without charge density wave (CDW) instability [17]. Since the $MA_2Z_4$ monolayers lacking inversion symmetry possess similar Zeeman-type spin-valley locking as the $MoS_2$-family monolayers [17,40-42], one might expect that if superconducting, they may have a high $B_c$. Lo and behold, we found that some superconducting 2D $MA_2Z_4$ monolayers have the highest $B_c/B_p$ ratio to date, to the best of our knowledge, because of cooperating of both IBCS and FFLO mechanisms.

Our discovery is partly enabled by our recent development of a first-principles computational approach for superconductivity [43,44], by self-consistently solving superconducting gap equation constructed from density-functional-theory based Wannier functions (WFs), especially in the presence of external magnetic field. It allows us to predict not only $T_c$ but also $B_c$ of a superconductor, as well as topological superconductors [43,44]. Applying this method, we have systematically investigated the field-dependent superconductivity of $\alpha_1$-$TaSi_2P_4$, $\alpha_1$-$TaSi_2N_4$, $\alpha_1$-$NbSi_2P_4$, $\alpha_2$-$TaGe_2P_4$, and $\alpha_2$-$NbGe_2P_4$ monolayers. We found that without magnetic field, $T_c$ is ~22.5 K for $\alpha_1$-$TaSi_2N_4$ and below 10 K for others. The superconductivity in $\alpha_1$-$TaSi_2P_4$, $\alpha_1$-$TaSi_2N_4$ and $\alpha_2$-$TaGe_2P_4$ sustains until the field reaches up to ~50 tesla, due to a large Zeeman field $B_{eff}$ induced by strong SOC of Ta. By fitting the normalized $T_c$ at low field using a microscopic model of Ising superconductor, $B_c^{Ising}$ is estimated to be ~30, ~20, ~10, ~50, ~20 times of $B_p$ at zero Kelvin (0 K) for $\alpha_1$-$TaSi_2P_4$, $\alpha_1$-$TaSi_2N_4$, $\alpha_1$-$NbSi_2P_4$, $\alpha_2$-$TaGe_2P_4$, and $\alpha_2$-$NbGe_2P_4$, respectively. The total $B_c$'s are shown to be further enhanced, as much as tripling the $B_c^{Ising}$, by the additional FFLO paring mechanism.

**Results and discussion**

The atomic structures of $MA_2Z_4$ monolayer can be viewed as the $MoS_2$-type $MZ_2$ monolayer with the surface dangling bonds passivated by InSe-type $A_2Z_2$, constituting a septuple layer of Z-A-Z-M-Z-A-Z [16,17]. This unique sandwich structure creates a large $MA_2Z_4$ family with diverse properties arising from varying compositions and relative positions between atomic planes. Here we focus on the nonmagnetic metal compounds which are stable in the $\alpha_1$ and $\alpha_2$ phases, including $\alpha_1$-$TaSi_2P_4$, $\alpha_1$-$TaSi_2N_4$, $\alpha_1$-$NbSi_2P_4$, $\alpha_2$-$TaGe_2P_4$, and $\alpha_2$-$NbGe_2P_4$ [17]. From Fig. 1a and 1b, one sees that these non-centrosymmetric phases lack inversion symmetry but contain out-of-plane mirror symmetry $m_z$. Consequently, SOC induces a $B_{eff}$ to orient electron spins in the out-of-plane direction, manifesting a Zeeman-type spin-valley locking [17,40-42]. This feature can be clearly seen from Fig. 1c and 1d (see also Extended Data Fig. 1) obtained from the first-principles calculations (see Methods for details). Moreover, the up- and down-spin branches cross with each other along the M-Γ-M' **k**-point paths, forming three $m_z$-protected Weyl nodal lines (see Section I of Supplementary Information, SI). The generic metallic nature and spin-valley locking provide the precondition for the intrinsic

IBCS pairing, while the 1D Weyl states are shown to initiate and stabilize the extrinsic FFLO pairing under magnetic field by lifting spin degeneracy [27,28,45].

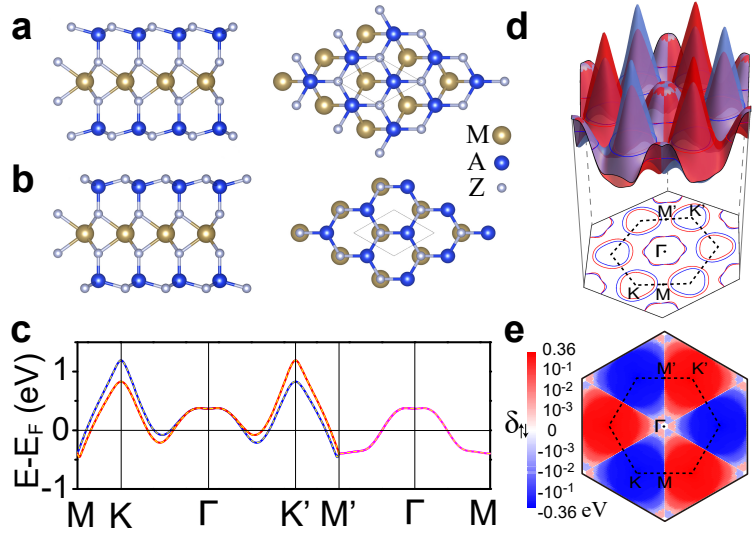

**Figure 1. The atomic and electronic structures. a**, **b**, Side (left) and top (right) views of (**a**) $\alpha_1$-MA$_2$Z$_4$ and (**b**) $\alpha_2$-MA$_2$Z$_4$ monolayer. **c,** Band structure of $\alpha_1$-TaSi$_2$P$_4$ monolayer, showing typical features of spin-valley locking. Red and blue lines denote respectively the split up- and down-spin states, while the pink line indicates the spin-degenerated Weyl nodal line. Yellow dashed lines are the WF fits of band structure. **d**, 3D band structure (top panel) and 2D FS contour (bottom) plots. **e**, Distribution of spin-splitting $\delta_{\uparrow\downarrow} = E_\uparrow - E_\downarrow$, i.e. the effective Zeeman field $B_{eff} = \delta_{\uparrow\downarrow}/\mu_\mathbf{B}$ for the two metallic bands, which vanishes along the M-Γ-M' paths (white lines).

To quantitatively characterize the anticipated superconductivity in the considered MA$_2$Z$_4$ monolayers, we first calculated the electron-phonon coupling (EPC) strength $\lambda$ and estimated critical temperature $T_\mathrm{c}^\mathrm{AD}$ using the Allen-Dynes modified McMillan's formula (see Methods for details). Phonon spectra calculations indicate that EPC induces a phonon mode softening, but no CDW instability which is known to be detrimental to IBCS [20-23] and FFLO pairing [45]. We summarize the superconductivity related parameters, e.g. EPC $\lambda$, logarithmically averaged frequency $\langle\omega\rangle_\mathrm{log}$, and $T_\mathrm{c}^\mathrm{AD}$, in Table 1. The convergence of these values was carefully checked with respect to different parameter settings (Extended Data Table 1) and pseudopotentials (Extended Data Table 2). Qualitatively, the finite $T_\mathrm{c}^\mathrm{AD}$ (Table 1) indicates that all five MA$_2$Z$_4$ monolayers are likely to exhibit intrinsic superconductivity without the need of doping.

**Table 1. Summary of the calculated parameters for superconductivity in the five $MA_2Z_4$ monolayers.**

|  | $\alpha_1$-TaSi$_2$P$_4$ | $\alpha_1$-TaSi$_2$N$_4$ | $\alpha_1$-NbSi$_2$P$_4$ | $\alpha_2$-TaGe$_2$P$_4$ | $\alpha_2$-NbGe$_2$P$_4$ |
|---|---|---|---|---|---|
| $\lambda$ | 0.77 | 1.29 | 0.79 | 0.66 | 0.80 |
| $N_F$ (eV$^{-1}$) | 2.04 | 1.92 | 2.22 | 1.96 | 2.18 |
| $\langle\omega\rangle_{\log}$ (K) | 105.56 | 231.17 | 136.93 | 105.48 | 115.25 |
| $T_c^{AD}$ (K) | 4.60 | 22.46 | 6.31 | 3.12 | 5.35 |
| $g$ | 0.28 | 0.57 (0.41) | 0.27 | 0.23 | 0.27 |
| $\Delta_0$ (meV) | 0.74 | 6.91 | 0.98 | 0.54 | 0.85 |
| $T_c^{SCF}$ (K) | 4.90 | 47.00 (22.50) | 6.45 | 3.55 | 5.60 |
| $B_c$ (tesla) | ~691 | ~1296 | ~190 | ~691 | ~241 |
| $B_p$ (tesla) | 9.04 | 42.08 | 11.98 | 6.55 | 10.37 |
| $\beta_{SOC}^*$ (meV) | 12.5 | 36.0 | 6.0 | 18.0 | 9.0 |

Since the above conventional methods cannot be applied to systems without time-reversal symmetry, we next further investigate the superconductivity and its dependence on magnetic field for the five $MA_2Z_4$ monolayers, by the newly developed method of WF construction of Bogoliubov-de Gennes (BdG) Hamiltonian $H_{BdG}^B(\mathbf{k})$ under magnetic field $B$ and self-consistently solving the resulting gap equation (see Methods for details). Figure 2a shows the calculated pairing gap $\Delta$ of $\alpha_1$-TaSi$_2$P$_4$ as a function of $T$ and $B$. Without magnetic field, the pairing gap at 0 K is $\Delta_0$~0.74 meV, which is fully suppressed at $T_c^{SCF}$~4.90 K, in good agreement with $T_c^{AD}$~4.60 K obtained from the Allen-Dynes-McMillan's formula. Due to Pauli paramagnetic pair-breaking effect, $T_c$ decreases with the increasing field. The gap is found to close at 0 K when $\mu_B B$ ($\mu_B$ is Bohr magneton) is larger than ~40 meV, translating to a $B_c$~691 tesla for $\alpha_1$-TaSi$_2$P$_4$ monolayer.

Similarly, $\Delta(T, B)$ is evaluated for $\alpha_1$-TaSi$_2$N$_4$, $\alpha_1$-NbSi$_2$P$_4$, $\alpha_2$-TaGe$_2$P$_4$, and $\alpha_2$-NbGe$_2$P$_4$ (Extended Data Fig. 4), using the parameters listed in Table 1. Overall, $T_c^{SCF}$ agrees well with $T_c^{AD}$ for those MA$_2$Z$_4$ monolayers with intermediate EPC strength ($\lambda$ < 1.0). But for $\alpha_1$-TaSi$_2$N$_4$ with $\lambda$=1.29, a case of strong EPC, ~47.00 K is overestimated by about two times in comparison with $T_c^{AD}$~22.46 K. This indicates that the first-principles WF gap equation can be safely applied to superconductors with intermediate EPC strength. For example, if $g$ were lowered to 0.41 for $\alpha_1$-TaSi$_2$N$_4$, its $T_c^{SCF}$ would be reduced to $T_c^{AD}$ (Table 1).

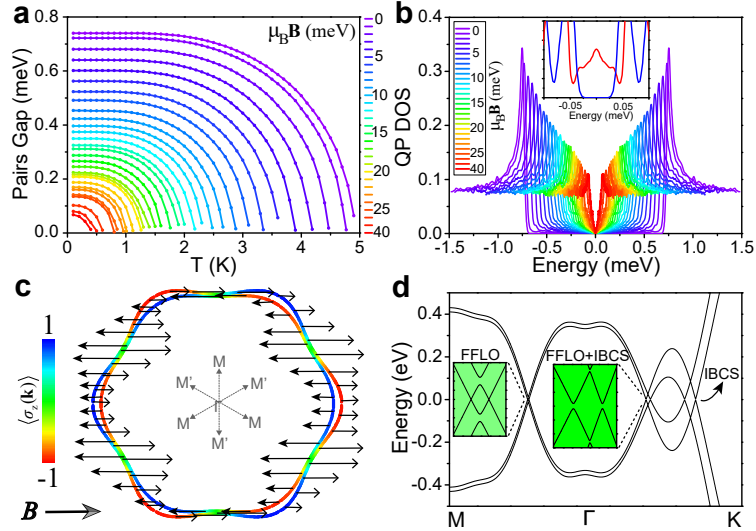

**Figure 2. IBCS and FFLO pairing of $\alpha_1$-TaSi$_2$P$_4$ monolayer. a**, The temperature-dependent pairing gaps under different in-plane magnetic fields (color bar). **b,** The dependence of QP DOS on the in-plane magnetic field at 0.1 K. Inset shows the magnified view of the QP DOS for $\mu_B B$=0.7 (blue line) and 0.8 (red) meV. **c**, Plots of spin texture on the FS contours centered at $\Gamma$ point with $\mu_B B$=10 meV. The colors in (c) denote the out-of-plane spin component, and the arrows denote the in-plane components. **d**, Plot of QP dispersion relation under the in-plane magnetic field. Insets show the magnified views of IBCS and/or FFLO pairs.

On the other hand, the estimated $B_c$ for $\alpha_1$-TaSi$_2$N$_4$, $\alpha_1$-NbSi$_2$P$_4$, $\alpha_2$-TaGe$_2$P$_4$, and $\alpha_2$-NbGe$_2$P$_4$ are 1296, 190, 691, and 241 tesla (Table 1), respectively, which are all higher than most known Ising superconductors. In order to better understand the physical origin of such a high critical field, we take $\alpha_1$-TaSi$_2$P$_4$ as an example to calculate and analyze the superconducting quasi-particle (QP) density of states (DOS) from the solutions of $H_{BdG}^{B}(\mathbf{k})$. When $B = 0$, the QP DOS is exactly zero within the pairing gap, showing a standard U-shape at all temperatures below $T_c^{SCF}$ (Extended Data Fig. 2g). This is because time-reversal-symmetry is preserved and spin orientations are exactly anti-parallel for the electrons at opposite momenta, which condense into ideal BCS pairs with *s*-wave symmetry. Differently, non-zero QP DOS emerges inside the pairing gap when $B \neq 0$, and the pairing gap gradually evolves into a V-shape (Fig. 2b). The energy window for zero QP DOS will completely close when $\mu_B B > 0.7$ meV (inset of Fig. 2b). This is reasonable when one considers the Pauli limit, as defined by $B_p = \Delta_0 / (\sqrt{2} \mu_B)$, is around ~9 tesla for $\alpha_1$-TaSi$_2$P$_4$ monolayer, which corresponds to $\mu_B B$ ~0.6 meV.

The above QP DOS behavior is fundamentally rooted in breaking of time-reversal-symmetry and the emergence of FFLO pairing in addition to IBCS pairing under magnetic field. To reveal this, we plot electron spin orientations under in-plane magnetic field in Fig. 2c (see also Extended Data Fig. 5). An important new feature, absent without magnetic field (lower panel of Fig. 1d), is that spins develop an in-plane component along the field direction, especially for the spins on the M-Γ-M' **k**-point paths where $B_{eff}$ is zero. The external magnetic field will change the eigenvalues of these states in an opposite manner to lift the degeneracy of Weyl nodal lines (Extended Data Fig. 5d). One immediate consequence is to make the BCS pairing no longer favored for these states when the field is close to or exceeds $B_p$. Instead, Cooper pairs with non-zero momentum, i.e. the FFLO pairs, start to form between electrons at different FS contours, especially around the Weyl points with weak local field $B_{eff}$ (Fig. 2c). So, the FFLO pairs form with "effectively 1D FS" along three Weyl nodal lines, and increase with the increasing magnetic field, as more electron spins in the vicinity will be re-oriented with those having zero out-of-plane component to satisfy the FS nesting condition (green color in Fig. 2c).

To further illustrate the simultaneous formation of both IBCS and FFLO pairs, we calculate the band dispersion of QP by diagonalizing $H_{BdG}^{B}(\mathbf{k})$. From Fig. 2d, one sees that there remain two nodal points in between as the magnetic field lifts the degeneracy elsewhere along the Γ-M (M') lines (Extended Data Fig. 5d). Since we did not consider the non-zero momentum **q** (see Eq. 9 in Mathods section), this exotic state corresponds to the so-called Sarma state featured with non-zero QP DOS (Fig. 3b) [46]. This homogeneous state is usually unstable with fixed chemical potentials [47,48], because it has the maximum Helmholtz free energy, which can be lowered by forming FFLO pairs with nonzero momentum **q** [49]. For convenience, we mark the Sarma state as "FFLO" in Fig.2d, from which the estimated $B_c$ should represent a lower limit because the actual FFLO pairing with a finite **q** is more stable. The evolution of QP dispersion from the Sarma to FFLO state can be shown by an effective model (see Section II of SI). Furthermore, both IBCS and FFLO pairs coexist along the Γ-K line on the side closer to Γ (i.e., FS around Γ) because of a moderate $B_{eff}$, while pure IBCS pairs prevail on the side closer to K (i.e., FS around K) because of a strong $B_{eff}$. These analyses indicate that both IBCS and FFLO pairing mechanisms self- and co-operate in different regions of Brillouin zone (BZ), resulting in an unprecedented high critical field $B_c$.

Now, let's further analyze the temperature dependence of $B_c(T)$. Figure 3a shows obviously an upturn of $B_c(T)$ when $T \to 0$ K, which is consistent with IBCS or FFLO pairing. Near $T_c^{SCF}$, in contrast to the orbital limited case with a finite slope, the dependence is known as $B_c \propto \sqrt{T_c^{SCF} - T}$ [20], which is clearly seen for $\alpha_1$-NbSi$_2$P$_4$ and $\alpha_2$-NbGe$_2$P$_4$ monolayer (inset of Fig. 3a). For $\alpha_1$-TaSi$_2$P$_4$, $\alpha_1$-TaSi$_2$N$_4$, and $\alpha_2$-TaGe$_2$P$_4$, the magnetic fields up to 50 tesla still show little suppressing effect on $T_c^{SCF}$, which can be attributed to a large $B_{eff}$ induced by strong SOC of Ta atom. To give a semi-quantitative measure on the robustness of superconductivity against the field, we evaluate the normalized critical field by Pauli limit, $B_c/B_p$ as a function of the normalized temperature, $T/T_c^{SCF}$, as shown in Fig. 3b. At 0 K, the $B_c/B_p$ ratio can reach as high as ~100 in $\alpha_2$-TaGe$_2$P$_4$, and even the lowest ratio in $\alpha_1$-NbSi$_2$P$_4$ is still larger than 10.

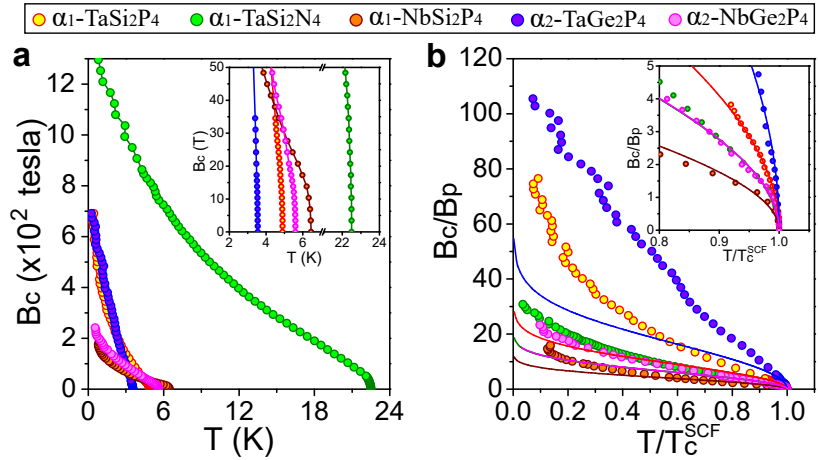

**Figure 3. The temperature $T$ dependent in-plane critical magnetic field $B_c$ for the five MA$_2$Z$_4$ monolayers. a, b,** The dependences of $B_c$ on $T$ (**a**) and the $B_c/B_p$ ratios as a function of normalized temperature $T/T_c^{SCF}$ (**b**). Insets show the magnified views under low magnetic field. The solid lines in (**b**) are the fits of the dependence of $B_c/B_p$ on $T/T_c^{SCF}$ by the microscopic model of Ising superconductor.

We emphasize again that the high $B_c/B_p$ ratio is enabled by cooperating of both IBCS and FFLO pairing mechanisms. In this regard, it will be useful to estimate roughly the contribution of IBCS pairing alone and hence their relative contribution, which is done by using a widely applied microscopic model of Ising SOC [18] (see Methods for details). Using the fitting parameter $\beta_{SOC}^*$ (Table 1), one can reproduce well the dependence of $B_c/B_p$ on $T/T_c^{SCF}$ under low magnetic field (see inset of Fig. 3b), when the IBCS pairing is dominant. With the increasing field, the deviation between the microscopic model and the self-consistent solution increases noticeably (see main figure of Fig. 4b), because more FFLO pairs start to form. At 0 K, $B_c^{Ising}/B_p$ is fit to be about 30, 20, 10, 50, and 20 for $\alpha_1$-TaSi$_2$P$_4$, $\alpha_1$-TaSi$_2$N$_4$, $\alpha_1$-NbSi$_2$P$_4$, $\alpha_2$-TaGe$_2$P$_4$, and $\alpha_2$-NbGe$_2$P$_4$, respectively. These high $B_c^{Ising}/B_p$ ratios are plausible to be realized in experiments even if the samples are not so clean.

It is worth noting that even without FFLO pairs, the $B_c^{Ising}/B_p$ ratios are still higher than most of the experimentally confirmed Ising superconductors except for WS$_2$ [24]. This is likely because the MA$_2$Z$_4$ monolayer neither possesses CDW instability nor needs ionic gating to evoke superconductivity. Due to the absence of CDW phase, the $B_c^{Ising}/B_p$ ~20 of WS$_2$ is known to be larger than that (~5) of NbSe$_2$ with stronger Ising SOC, while the gating induced Rashba SOC causes a downturn of the ratio at low temperature [24]. As a benchmark for our prediction, we have self-consistently solved the $\Delta$ ($T$, $B$) of electron-doped WS$_2$ [24]. Our results reproduce reasonably well the experimental $B_c/B_p$ (see Section III of SI), which confirms the reliability of our predictions. If clean MA$_2$Z$_4$ monolayer is to be synthesized, intrinsic superconductivity is expected to be observed, as previously done for NbS$_2$ related materials [31,32], and most appealingly, unprecedented high $B_c$ that can triple $B_c^{Ising}$ (Extended Data Fig. 6) as well as the unusual cooperating IBCS and FFLO pairing mechanisms may be experimentally revealed. Finally, exploring possible topological superconducting state associated with the Weyl nodal lines, such as the one associated with the Sarma state [50], in these Ising superconductors could be an interesting topic of future study.

**Online content**

Any methods, additional references, extended data, supplementary information, acknowledgements, peer review information, details of author contributions, competing interests, and statements of data and codes are available at https.

**Methods**

**Details of First-principles calculations**

The QUANTUM ESPRESSO (QE) package [51] was used to calculate the electronic property and phonon spectra as well as the electron-phonon coupling (EPC) strength based on the density functional theory (DFT) and the density-functional perturbation theory (DFPT) [52], and the projector-augmented wave (PAW) datasets with the functional type of PBE [53] was employed and the kinetic energy cutoff was set to 60 Ry for wave functions. The atomic structures of the considered five $MA_2Z_4$ monolayers were set up by introducing a vacuum region of more than 15 Å to avoid the interactions between neighboring periodic images. Structural relaxations and self-consistent calculations were performed on a uniform 16×16×1 **k**-point sampling in the first BZ. The dynamic matrix and phonon frequency are computed on the 8×8×1 **q**-points mesh with the 16×16×1 **k**-points sampling, and a finer 32×32×1 **k**-point grid is used for the EPC calculations. Other **q**/**k**-point samplings (Extended Data Table 1) and the ultrasoft pseudopotentials (Extended Data Table 2) are employed to check the convergence of EPC calculations. The phonon DOS $F(\omega)$ and the isotropic Eliashberg spectral function $\alpha^2F(\omega)$ as well as the cumulative frequency-dependent EPC strength $\lambda(\omega)$ are calculated using a 100×100×1 **q**-point sampling by means of the Fourier interpolation. The Wannier functions (WFs) are obtained by interfacing with the WANNIER90-2.1 code [54], where a single virtual orbital (e.g. *s* orbital) localized at the site of M atom was employed as the initial guess for the unitary transformations without minimization procedure and the **k**-points sampling is 60×60×1 in the first BZ.

**EPC calculations**

In the EPC calculations, the **q**- and *v*-resolved EPC strength $\lambda_{\mathbf{q}v}$ is given by:

$$\lambda_{\mathbf{q}v} = \frac{1}{N_F \omega_{\mathbf{q}v}} \sum_{mn,\mathbf{k}} W_{\mathbf{k}} |g_{mn,v}(\mathbf{k},\mathbf{q})|^2 \delta(\varepsilon_{n\mathbf{k}})\delta(\varepsilon_{m\mathbf{k}+\mathbf{q}}), \quad (1)$$

$$g_{mn,v}(\mathbf{k},\mathbf{q}) = \left(\frac{\hbar}{2M_0\omega_{\mathbf{q}v}}\right)^{1/2} \langle \psi_{m\mathbf{k}+\mathbf{q}} | \partial_{\mathbf{q}v}\Xi | \psi_{n\mathbf{k}} \rangle, \quad (2)$$

where $N_F$ is the electronic DOS at the Fermi level, $W_{\mathbf{k}}$ is the weight of wavevector **k**, $\varepsilon_{n\mathbf{k}}$ is the eigenvalue for electronic wavefunction $\psi_{n\mathbf{k}}$ with band index *n* and wavevector **k**, $\omega_{\mathbf{q}v}$ is the frequency of a phonon branch *v* at wavevector **q**, $\hbar$ is the reduced Planck constant, and $M_0$ is the ionic mass. $g_{mn,v}(\mathbf{k},\mathbf{q})$ represents the scattering amplitude between the electronic states with wavefunction $\psi_{n\mathbf{k}}$ and $\psi_{m\mathbf{k}+\mathbf{q}}$, induced by derivative ($\partial_{\mathbf{q}v}\Xi$) of the self-consistent potential associated with phonon $\omega_{\mathbf{q}v}$. $\delta$ is the Dirac delta function. The frequency $\omega$-dependent isotropic Eliashberg spectral function $\alpha^2F(\omega)$ and the cumulative EPC strength $\lambda(\omega)$ are then calculated as:

$$\alpha^2F(\omega) = \frac{1}{2}\sum_{\mathbf{q}v} W_{\mathbf{q}} \omega_{\mathbf{q}v} \lambda_{\mathbf{q}v} \delta(\omega - \omega_{\mathbf{q}v}), \quad (3)$$

$$\lambda(\omega) = 2\int_0^\omega d\omega' \frac{\alpha^2F(\omega')}{\omega'}, \quad (4)$$

Here $W_{\mathbf{q}}$ is the weight of wavevector **q**. The total EPC constant $\lambda$ equals to $\lambda(\omega_{max})$ with $\omega_{max}$ being the maximum of phonon frequency.

Specifically, taking $\alpha_1$-TaSi$_2$P$_4$ as an example, the calculated phonon spectra (Extended Data Fig. 2a) indicates no CDW instability, which is known to be detrimental to IBCS [20-23] and FFLO pairing [45]. But there exists a phonon softening mode along the Γ-M direction, which comes from the in-plane vibrations of Ta atom (Extended Data Fig. 2b) and couples strongly with electrons. This is because the metallic states around the Fermi level stem mainly from the $d_{xy}$, $d_{x^2-y^2}$, and $d_{z^2}$ orbitals of Ta atom (Extended Data Fig. 2c, d), which hybridize to form a 2D electron gas distributed symmetrically about the middle Ta atomic plane, as shown in Extended Data Fig. 2e. It is noted that these orbitals also form other states away from the Fermi level. The total EPC $\lambda$ is calculated to be 0.77 from the cumulative EPC $\lambda(\omega)$, and the logarithmically averaged frequency $\langle\omega\rangle_{\log}$ is evaluated to be 105.56 K from the Eliashberg spectral function $\alpha^2F(\omega)$ (Extended Data Fig. 2f). The key features of $\alpha_1$-TaSi$_2$P$_4$ are also found in other MA$_2$Z$_4$ monolayers (Extended Data Fig. 3), namely the absence of CDW instability and the EPC induced phonon softening, whose EPC $\lambda$ and $\langle\omega\rangle_{\log}$ are summarized in Table 1.

**Estimating critical temperature**

We applied two methods to estimate the superconducting critical temperature of the five considered MA$_2$Z$_4$ monolayers. One is based on the widely used conventional Allen-Dynes modified McMillan's formula [55,56]:

$$T_c^{AD} = \frac{\langle\omega\rangle_{\log}}{1.2}\exp\left[\frac{-1.04(1+\lambda)}{\lambda-\mu^*(1+0.62\lambda)}\right], \quad (5)$$

where the logarithmically averaged frequency $\langle\omega\rangle_{\log}$ is defined as:

$$\langle\omega\rangle_{\log} = \exp\left[\frac{2}{\lambda}\int\alpha^2F(\omega)\log(\omega)\frac{d\omega}{\omega}\right], \quad (6)$$

The $\mu^*$ represents the retarded Coulomb pseudopotential and has the following relationship with the screened Coulomb potential $\mu$ [57]:

$$\mu^* = \frac{\mu}{1+\mu\ln(\varepsilon_F/\theta_D)}, \quad (7)$$

where $\varepsilon_F$ and $\theta_D$ is the characteristic electron energy and Debye temperature, respectively. Using the typical value of $\mu^* = 0.1$, we estimated the $T_c^{AD}$ of the considered MA$_2$Z$_4$ monolayers (see Table 1). Together with the values of EPC $\lambda$ and $\langle\omega\rangle_{\log}$, the convergence of $T_c^{AD}$ was carefully checked with respect to different parameter settings (Extended Data Table 1). Our estimated ~3.12 K for $\alpha_2$-TaGe$_2$P$_4$ is reasionablly consistent with the previous report (3.75 K), but that for $\alpha_1$-TaSi$_2$N$_4$ (~22.46 K) is notably higher than the reported value (9.67 K) [17]. We thus further checked the results of $\alpha_1$-TaSi$_2$N$_4$ using different pseudopotentials (Extended Data Table 2), and our results remain self-consistent. Qualitatively, the finite $T_c^{AD}$ (Table 1) indicates that all these MA$_2$Z$_4$ monolayers hold high possibility of exhibiting intrinsic superconductivity without the need of doping.

The other one, $T_c^{SCF}$, is estimated from our recently developed first-principles approach based on WFs [43,44], by self-consistently solving the quasi-particle gap equation without and with external magnetic field. Important to the present study, it allows one to calculate not only the critical temperature $T_c^{SCF}$ but also the critical magnetic field $B_c$, namely the field under which the pairing gap closes. In general, there is a good agreement between $T_c^{AD}$ and $T_c^{SCF}$ at zero magnetic field when $\lambda < 1.0$, and $T_c^{SCF}$ tends to be overestimated in comparison with $T_c^{AD}$ when $\lambda > 1.0$ (Table 1).

Specifically, because the one metallic band of our particular interest after spin splitting is well isolated from others (Fig. 1c and Extended Data Fig. 1), we opt to use a single virtual orbital localized at the site of M atom as the initial guess to construct the WFs [58]. The resulting Hamiltonian $h_{\text{WFs}}(\mathbf{k})$ can reproduce very well the first-principles band structures under the basis of $\varphi_{\text{WFs}}^{\mathbf{k}} = (\varphi_{\mathbf{k}\uparrow}, \varphi_{\mathbf{k}\downarrow})^T$ (Fig. 1c). Then the BdG Hamiltonian $H_{\text{BdG}}^B(\mathbf{k})$ including an in-plane magnetic field $B$ along $x$ direction is constructed under the basis of $\psi_{\mathbf{k}} = (\varphi_{\mathbf{k}\uparrow}, \varphi_{\mathbf{k}\downarrow}, \varphi_{-\mathbf{k}\uparrow}^\dagger, \varphi_{-\mathbf{k}\downarrow}^\dagger)^T$:

$$H_{\text{BdG}}^B = \sum_{\mathbf{k}} \psi_{\mathbf{k}}^\dagger H_{\text{BdG}}^B(\mathbf{k}) \psi_{\mathbf{k}} \quad , \quad (8)$$

$$H_{\text{BdG}}^B(\mathbf{k}) = \begin{pmatrix} h_{\text{WFs}}(\mathbf{k}) + \mu_B B \sigma_x & i\Delta\sigma_y \\ -i\Delta\sigma_y & -[h_{\text{WFs}}(-\mathbf{k}) + \mu_B B \sigma_x]^* \end{pmatrix} \quad , \quad (9)$$

where $\mu_B$ is the Bohr magneton and $\sigma_{x/y}$ are Pauli matrices. The pairing gap $\Delta$ is assumed to be real, which is solved self-consistently using the following gap equation [43,44]:

$$\Delta = -\frac{g}{2V}\sum_{l,\mathbf{k}>0}\frac{1}{1+e^{\beta E_{l,\mathbf{k}}}}\frac{\partial E_{l,\mathbf{k}}}{\partial \Delta}\ ,\quad (10)$$

$\beta = \frac{1}{k_B T}$, $k_B$ is the Boltzmann constant and $T$ is temperature. The pairing strength is defined as $g = (\lambda-\mu)/N_F$ [59]. $\mu$ is the screened Coulomb potential, which is related to but usually larger than $\mu^*$ (Eq. 7). $N_F$ is the electron DOS at the Fermi level. $V$ is volume. $E_{l,\mathbf{k}}$ is the eigenvalue of $H_{\mathrm{BdG}}^{B}(\mathbf{k})$ with band index $l$ at $\mathbf{k}$ point. The summation is performed in half of the BZ ($\mathbf{k}>0$) due to particle-hole symmetry and only the quasi-particle states within the logarithmically averaged frequency, $|E_{l,\mathbf{k}}| \leq \langle\omega\rangle_{\log}$, are included. The values of these parameters used for self-consistently estimating the $T$ and $B$ dependent $\Delta$ of the five considered $MA_2Z_4$ monolayers (Fig. 2a and Extended Data Fig. 4) are listed in Table 1.

**Microscopic model of Ising superconductors**

The $B_c/B_p$ as a function of $T/T_c^{\mathrm{SCF}}$ at low fields (inset of Fig. 4b) was fitted by employing the microscopic model with only Ising SOC [16]:

$$\ln\left(\frac{T}{T_c^{\mathrm{SCF}}}\right) + \Phi(\rho_-) + \Phi(\rho_+) + [\Phi(\rho_-) - \Phi(\rho_+)]\frac{\boldsymbol{\beta}_{\mathrm{SOC}}^2 - \mathbf{b}^2}{|\boldsymbol{\beta}_{\mathrm{SOC}} - \mathbf{b}||\boldsymbol{\beta}_{\mathrm{SOC}} + \mathbf{b}|} = 0\ ,\quad (11)$$

where $\rho_\pm \equiv \frac{|\boldsymbol{\beta}_{\mathrm{SOC}} + \mathbf{b}| \pm |\boldsymbol{\beta}_{\mathrm{SOC}} - \mathbf{b}|}{2\pi k_B T_c}$ with $\mathbf{b} = (\mu_B B_c^{\mathrm{Ising}}, 0, 0)$ and $\boldsymbol{\beta}_{\mathrm{SOC}} = (0, 0, \beta_{\mathrm{SOC}}^*)$, and $\beta_{\mathrm{SOC}}^*$ is the effective Ising SOC strength. $\Phi(\rho)$ is related to the digamma function $\psi(X)$:

$$\Phi(\rho) \equiv \frac{1}{2}\mathrm{Re}\left[\psi\left(\frac{1+i\rho}{2}\right) - \psi\left(\frac{1}{2}\right)\right]\ ,\quad (12)$$

This model is widely applied to fit the dependence of $B_c/B_p$ on $T/T_c^{\mathrm{SCF}}$ in the experimental studies of Ising superconductors.

**Data availability**

All data needed to evaluate the conclusions of this paper are available within the paper and Supplementary Information.

**Code availability**

The central code used in this paper is the QE and WANNIER90 code. Detailed informations related to the license and user guide are available at https://www.quantum-espresso.org/ and http://www.wannier.org/, respectively.

**Methods references**

**ACKNOWLEDGEMENTS**

F.L. acknowledges financial support from DOE-BES (No. DE-FG02-04ER46148). X.Z. acknowledges financial support by the National Natural Science Foundation of China (No. 12004357), the Natural Science Foundation of Shandong Province (No. ZR2020QA053), and the Young Talents Project at Ocean University of China.

**AUTHOR CONTRIBUTIONS**

X.Z. and F.L. conceived the research project. X.Z. performed the calculations. X.Z. and F.L. disccused results and wrote the manuscript.

## COMPETING INTERESTS

The authors declare no competing interests.

## ADDITIONAL INFORMATION

**Supplementary information** is available for this paper.
**Correspondence** and requests for materials should be addressed to F.L. and X.Z.
**Peer review information**
**Reprints and permissions information** is available at www.nature.com/reprints.

**Extended data figure/table legends**

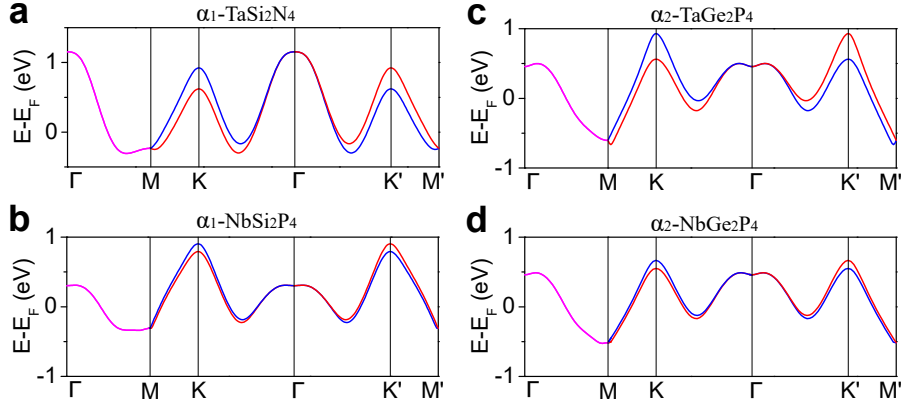

**Extended Data Fig. 1. Band structure with Weyl nodal lines and spin-valley locking. a-d**, The metallic bands of (**a**) $\alpha_1$-TaSi$_2$N$_4$, (**b**) $\alpha_1$-NbSi$_2$P$_4$, (**c**) $\alpha_2$-TaGe$_2$P$_4$, and (**d**) $\alpha_2$-NbGe$_2$P$_4$. The pink line represents the doubly degenerated Weyl nodal lines, while the red and blue lines denote the up- and down-spin states, respectively. The Fermi level is set to the zero energy.

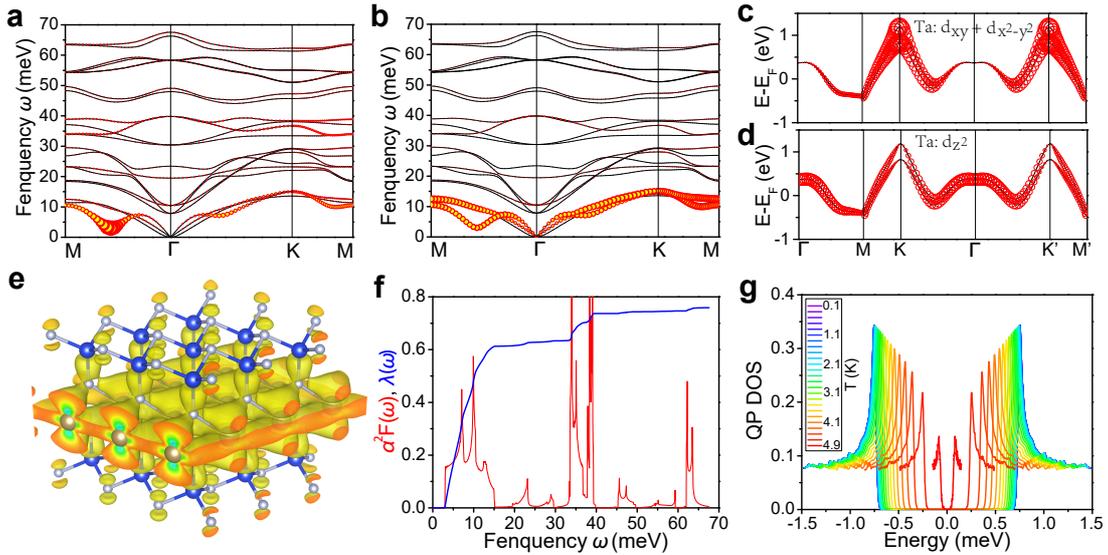

**Extended Data Fig. 2. EPC analysis of $\alpha_1$-TaSi$_2$P$_4$ monolayer. a**, **b**, Phonon spectra with the magnitude of (**a**) the EPC strength $\lambda_{\mathbf{q}\nu}$ and (**b**) the in-plane vibrations of Ta atom being drawn proportional to the size of yellow filled circles. **c**, **d**, The band structure with the contributions of (**c**) $d_{xy}+d_{x^2-y^2}$ and (**d**) $d_{z^2}$ orbitals of Ta atoms being indicated by red circles. **e**, The charge density distribution of the two isolated metallic bands of $\alpha_1$-TaSi$_2$P$_4$ monolayer. **f**, Plots of Eliashberg spectral function $\alpha^2 F(\omega)$ and cumulative frequency-dependent EPC strength $\lambda(\omega)$. **g**, The dependence of QP DOS on temperature without in-plane magnetic field.

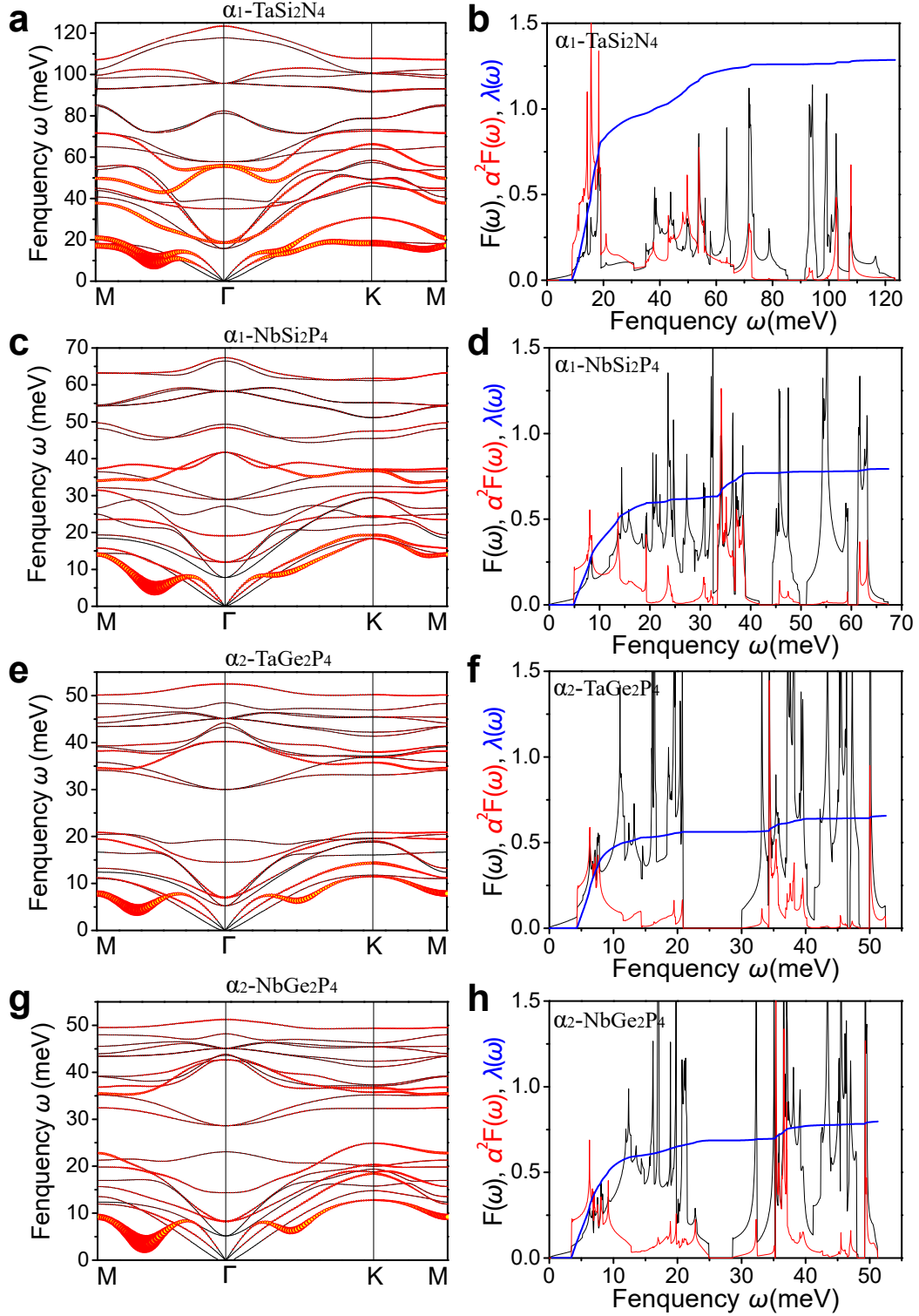

**Extended Data Fig. 3. EPC calculations of MA$_2$Z$_4$ monolayers. a**, **c**, **e**, **g**, The phonon spectra of (**a**) $\alpha_1$-TaSi$_2$N$_4$, (**c**) $\alpha_1$-NbSi$_2$P$_4$, (**e**) $\alpha_2$-TaGe$_2$P$_4$, and (**g**) $\alpha_2$-NbGe$_2$P$_4$ monolayer with the magnitude of EPC strength $\lambda_{\mathbf{q}\nu}$ being drawn proportional to the size of yellow filled circles. **b**, **d**, **f**, **h**, The plots of phonon DOS $F(\omega)$, Eliashberg spectral function $\alpha^2 F(\omega)$, and cumulative frequency-dependent EPC strength $\lambda(\omega)$ of (**b**) $\alpha_1$-TaSi$_2$N$_4$, (**d**) $\alpha_1$-NbSi$_2$P$_4$, (**f**) $\alpha_2$-TaGe$_2$P$_4$, and (**h**) $\alpha_2$-NbGe$_2$P$_4$ monolayer.

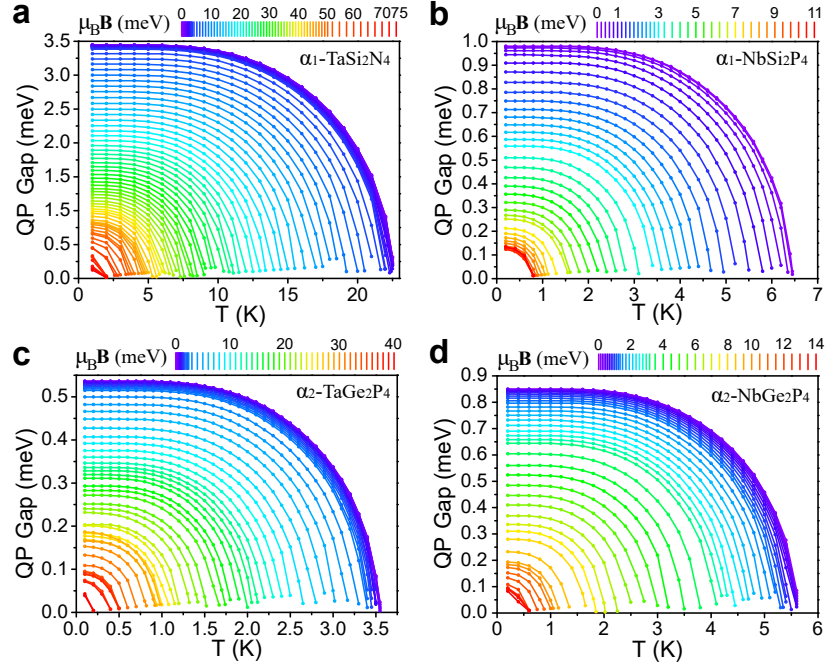

**Extended Data Fig. 4. The self-consistently calculated pairing gap at different temperature and in-plane magnetic field.** The temperature-dependent pairing gap of (a) $\alpha_1$-TaSi$_2$N$_4$, (b) $\alpha_1$-NbSi$_2$P$_4$, (c) $\alpha_2$-TaGe$_2$P$_4$, and (d) $\alpha_2$-NbGe$_2$P$_4$ monolayer under different in-plane magnetic field.

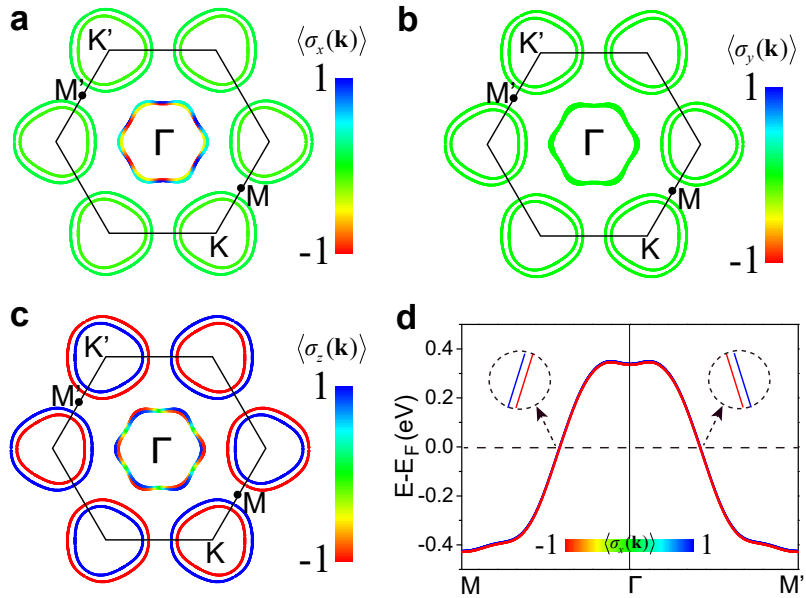

**Extended Data Fig. 5. The electron spin directions for the states on the FS of $\alpha_1$-TaSi$_2$P$_4$ monolayer under magnetic field $\mu_B B$=10 meV along the $x$ direction. a-c**, The expectation values of Pauli matrix (a) $\sigma_x$, (b) $\sigma_y$, and (c) $\sigma_z$. **d**, The bands along the M-Γ-M' k-point path, which are split by the in-plane magnetic field. The spins develop a component in parallel to the field direction.

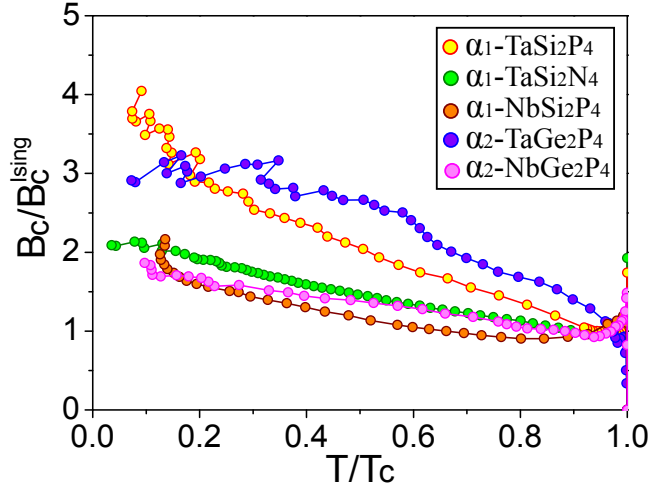

**Extended Data Fig. 6.** The ratio of $B_c/B_c^{Ising}$ versus $B_c/B_c^{Ising}$ for the five MA$_2$Z$_4$ monolayers.

**Extended Data Table 1.** The convergence of EPC $\lambda$, $\langle\omega\rangle_{\log}$, and $T_c^{AD}$ calculated by using different k- and q-mesh for the MA$_2$Z$_4$ monolayers.

|  | Fine k-mesh | Coarse k-mesh | q-mesh | EPC $\lambda$ | $\langle\omega\rangle_{\log}$ (K) | $T_c^{AD}$ (K) |
|---|---|---|---|---|---|---|
| α$_1$-TaSi$_2$P$_4$ | 32×32×1 | 16×16×1 | 4×4×1 | 1.25 | 68.96 | 6.48 |
|  | 36×36×1 | 18×18×1 | 6×6×1 | 0.67 | 125.07 | 3.96 |
|  | 32×32×1 | 16×16×1 | 8×8×1 | 0.77 | 105.56 | 4.60 |
| α$_1$-TaSi$_2$N$_4$ | 32×32×1 | 16×16×1 | 4×4×1 | 1.90 | 177.89 | 24.67 |
|  | 36×36×1 | 18×18×1 | 6×6×1 | 1.19 | 249.32 | 22.16 |
|  | 32×32×1 | 16×16×1 | 8×8×1 | 1.29 | 231.17 | 22.46 |
| α$_1$-NbSi$_2$P$_4$ | 32×32×1 | 16×16×1 | 4×4×1 | 1.23 | 93.82 | 8.65 |
|  | 36×36×1 | 18×18×1 | 6×6×1 | 0.73 | 153.89 | 5.94 |
|  | 32×32×1 | 16×16×1 | 8×8×1 | 0.79 | 136.93 | 6.31 |
| α$_2$-TaGe$_2$P$_4$ | 32×32×1 | 16×16×1 | 4×4×1 | 0.53 | 127.33 | 1.94 |
|  | 36×36×1 | 18×18×1 | 6×6×1 | 1.29 | 52.99 | 5.17 |
|  | 32×32×1 | 16×16×1 | 8×8×1 | 0.66 | 105.48 | 3.12 |
| α$_2$-NbGe$_2$P$_4$ | 32×32×1 | 16×16×1 | 4×4×1 | 0.82 | 110.29 | 5.42 |
|  | 32×32×1 | 16×16×1 | 8×8×1 | 0.80 | 115.25 | 5.35 |

**Extended Data Table 2.** The convergence of EPC $\lambda$, $\langle\omega\rangle_{\log}$, and $T_c^{AD}$ for the $\alpha_1$-TaSi$_2$N$_4$ monolayer by using ultrasoft pseudopotentials under different k- and q-mesh.

|  | Fine k-mesh | Coarse k-mesh | q-mesh | EPC $\lambda$ | $\langle\omega\rangle_{\log}$ (K) | $T_c^{AD}$ (K) |
|---|---|---|---|---|---|---|
| | 32×32×1 | 16×16×1 | 4×4×1 | 1.93 | 178.62 | 25.02 |
| $\alpha_1$-TaSi$_2$N$_4$ | 36×36×1 | 18×18×1 | 6×6×1 | 1.19 | 254.67 | 22.48 |
| | 32×32×1 | 16×16×1 | 8×8×1 | 1.30 | 229.35 | 22.54 |
| | 36×36×1 | 18×18×1 | 9×9×1 | 1.24 | 239.17 | 22.19 |

Supplementary information for

# Weyl nodal line induced pairing in Ising superconductor and high critical field

Xiaoming Zhang[1,*] and Feng Liu[2,*]

[1] *Department of Physics, College of Information Science and Engineering, Ocean University of China, Qingdao, Shandong 266100, China*

[2] *Department of Materials Science and Engineering, University of Utah, Salt Lake City, Utah 84112, USA*

*Correspondence to: fliu@eng.utah.edu, zxm@ouc.edu.cn

**Table of contents**

## Section I. Symmetry analyses of the Weyl nodal lines

The five considered monolayer $MA_2Z_4$ compounds are all non-centrosymmetric systems that lack the inversion symmetry but possess the out-of-plane mirror symmetry $m_z = i\tau_x\sigma_z$, the time-reversal symmetry, and the $C_{3z}$ rotational symmetry. Following a recently theoretical analysis on the systems with these symmetries [1], one can write an effective lattice Hamiltonian, up to the second order of $k$, as:

$$h(\mathbf{k}) = c_0\tau_0\sigma_0 + c_1\tau_x\sigma_0 + c_2\tau_y\sigma_z + c_3\tau_z\sigma_0 + c_4\tau_0(k_x\sigma_x + k_y\sigma_y) + c_5\tau_0(-k_y\sigma_x + k_x\sigma_y)$$
$$+ c_6\tau_x(k_x\sigma_x + k_y\sigma_y) + c_7\tau_x(-k_y\sigma_x + k_x\sigma_y) + c_8\tau_z(k_x\sigma_x + k_y\sigma_y)$$
$$+ c_9\tau_z(k_x\sigma_y - k_y\sigma_x) + c_{10}\tau_0\sigma_0 k^2 + c_{11}\tau_x\sigma_0 k^2 + c_{12}\tau_y\left[(k_x^2 - k_y^2)\sigma_x - 2k_xk_y\sigma_y\right]$$
$$+ c_{13}\tau_y\left[(k_x^2 - k_y^2)\sigma_y + 2k_xk_y\sigma_x\right] + c_{14}\tau_y\sigma_z k^2 + c_{15}\tau_z\sigma_0 k^2$$
, (S1)

where the Pauli matrices $\tau$ and $\sigma$ describe the mirrored plane and spin degrees of freedom. The $c_i$ ($i$=0,1,2,...,15) are constants and closely related to symmetry operations. For example, under the out-of-plane mirror symmetry $m_z$, the $c_i$ ($i$=2,3,4,5,6,7,14,15) are forbidden (Table S1) in the absence of inversion symmetry [1], which reduces the $h(\mathbf{k})$ to:

$$H = (c_0 + c_{10}k^2)\tau_0\sigma_0 + (c_1 + c_{11}k^2)\tau_0\sigma_x + c_8\tau_z(k_x\sigma_x + k_y\sigma_y) + c_9\tau_z(k_x\sigma_y - k_y\sigma_x)$$
$$+ c_{12}\tau_y\left[(k_x^2 - k_y^2)\sigma_x - 2k_xk_y\sigma_y\right] + c_{13}\tau_y\left[(k_x^2 - k_y^2)\sigma_y + 2k_xk_y\sigma_x\right]$$
, (S2)

The corresponding eigenvalues are:

$$E_{1,2} = c_0 + c_{10}k^2 - \sqrt{\begin{array}{c}(c_1 + c_{11}k^2)^2 + k^2(c_8^2 + c_9^2) + k^4(c_{12}^2 + c_{13}^2) \\ \pm\left[2k_x(k_x^2 - 3k_y^2)(c_9c_{12} - c_8c_{13}) - 2k_y(k_y^2 - 3k_x^2)(c_9c_{13} + c_8c_{12})\right]\end{array}}, \quad (S3)$$

$$E_{3,4} = c_0 + c_{10}k^2 + \sqrt{\begin{array}{c}(c_1 + c_{11}k^2)^2 + k^2(c_8^2 + c_9^2) + k^4(c_{12}^2 + c_{13}^2) \\ \pm\left[2k_x(k_x^2 - 3k_y^2)(c_9c_{12} - c_8c_{13}) - 2k_y(k_y^2 - 3k_x^2)(c_9c_{13} + c_8c_{12})\right]\end{array}}, \quad (S4)$$

One can easily identify that $E_1$ and $E_2$ (or $E_3$ and $E_4$) are doubly degenerate at the k-points satisfying $k_x(k_x^2 - 3k_y^2) = 0$ and $k_y(k_y^2 - 3k_x^2) = 0$, which constitute the Weyl nodal lines along the M'-Γ-M k-point paths in the Brillouin zone (see Fig. 1c-e).

**Table S1.** Transformation table giving the terms of $c_i$ ($i$=0,1,2,...,15), which are allowed (+) or not allowed (−) under out-of-plane mirror symmetry $m_z$ operation.

|  | $c_0$ | $c_1$ | $c_2$ | $c_3$ | $c_4$ | $c_5$ | $c_6$ | $c_7$ | $c_8$ | $c_9$ | $c_{10}$ | $c_{11}$ | $c_{12}$ | $c_{13}$ | $c_{14}$ | $c_{15}$ |
|---|---|---|---|---|---|---|---|---|---|---|---|---|---|---|---|---|
| $m_z$ | + | + | − | − | − | − | − | − | + | + | + | + | + | + | − | − |

## Section II. Evaluation of quasi-particles dispersion under magnetic field

To reveal the magnetic field induced evolution of superconducting quasi-particles (QP) dispersions, we adopt a single-particle Hamiltonian with the basis of $\phi_k = (\phi_{k\uparrow}, \phi_{k\downarrow})^T$, to describe a simple metallic band under the field $\mu_B B$ ($\mu_B$ is the Bohr magneton) along z direction:

$$H = \sum_k \phi_k^\dagger H(k) \phi_k, \quad (S5)$$

$$H(k) = \left(\frac{k^2}{2m} - \mu_0\right) I_{2\times 2} + \mu_B B \sigma_z \text{ with } k^2 = \sum_{i=1}^d k_i^2, \quad (S6)$$

where $\mu_0$ is the chemical potential that defines the Fermi level; $m$ is the effective mass and $d$ is the spatial dimension; $I_{2\times 2}$ and $\sigma_z$ (and the $\sigma_y$ below) are respectively the identity and Pauli matrix is spin space. Then the BdG Hamiltonian of FFLO pairing, or FF pairing exactly, with non-zero total momentum $q$ under the basis of $\chi_k \equiv (\phi_{k+q/2\uparrow}, \phi_{k+q/2\downarrow}, \phi^\dagger_{-k+q/2\uparrow}, \phi^\dagger_{-k+q/2\downarrow})^T$ is:

$$H_{BdG} = \frac{1}{2}\sum_k \chi_k^\dagger H_{BdG}(k) \chi_k, \quad (S7)$$

$$H_{BdG}(k) = \begin{pmatrix} H(k+q/2) & i\Delta\sigma_y \\ -i\Delta\sigma_y & -H^*(-k+q/2) \end{pmatrix}, \quad (S8)$$

Here $\Delta$ is a real number representing the pairing gap. By diagonalizing $H_{BdG}(k)$, one can easily obtain the energy spectra of QP:

$$E_\pm^+(k,q) = \frac{2kq + \sqrt{4\Delta^2 m^2 + (k^2+q^2-2m\mu_0)^2} \pm 2m\mu_B B}{2m}, \quad (S9)$$

$$E_\pm^-(k,q) = \frac{2kq - \sqrt{4\Delta^2 m^2 + (k^2+q^2-2m\mu_0)^2} \pm 2m\mu_B B}{2m}, \quad (S10)$$

They indicate that both the electron branch $E_\pm^+(k,q)$ and hole branch $E_\pm^-(k,q)$ are split by the magnetic field, while the non-zero $q$ will lead to unequal QP states at $k$ and $-k$.

Specifically, we plot the QP spectra under different external magnetic field and $q$ values in Fig. S1. Without field, the homogeneous BCS pairs with $q = 0$ are formed (Fig. S1a), with both the electron and hole branches being doubly degenerate. The degeneracy will be lifted by magnetic field $\mu_B B$, inducing a phase translation from the BCS (Fig. S1a and S1b) to the Sarma (Fig. S1d) state at $\mu_B B = \Delta$, featured with gapless excitations (Fig. S1c). The Sarma state (Fig. S1d), possessing nodal points at zero energy, is usually unstable for fixed chemical potentials due to its maximum Helmholtz free energy [2-4]. The Helmholtz free energy can be decreased by forming pairs with nonzero momentum $q$ (Fig. S1e-1h) [5], and reaches the minimum when $|q| = |k_{F\uparrow} - k_{F\downarrow}| = \left|\sqrt{2m(\mu_0 + \mu_B B)} - \sqrt{2m(\mu_0 - \mu_B B)}\right|$, corresponding to the formation of FFLO pairs (Fig. S1g). Here the $k_{F\uparrow/\downarrow}$ is the Fermi-surface momentum of spin $\uparrow/\downarrow$. The inhomogeneous nature of the FFLO state in the momentum space can be understood by the coexistence of unpaired states and the paired states with the middle of pairing gap $\Delta$ locating exactly at the zero energy, which is beneficial to maintain the superconductivity under the magnetic field beyond the Pauli paramagnetic limit.

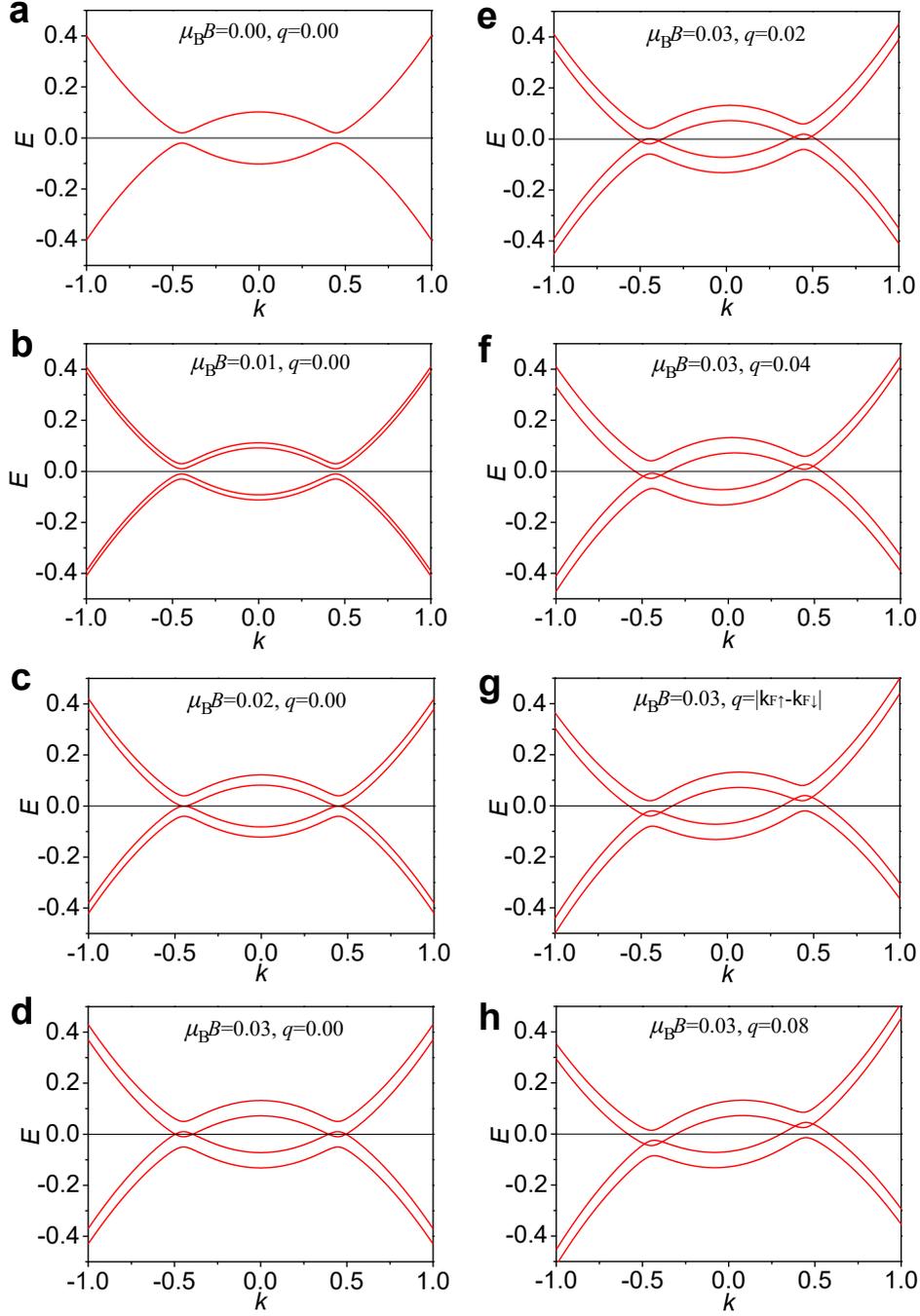

**Figure S1. Plot of QP dispersion relation using $m=1.0$, $\mu_0=0.1$, and $\Delta=0.02$ in the single-particle Hamiltonian.** The used values of $\mu_B B$ and $q$ are marked in each figure.

**Section III. Reproducing the superconductivity of electron-doped WS$_2$ monolayer under magnetic field**

As a benchmark for our predictions on the critical magnetic field of the MA$_2$Z$_4$ monolayers, we self-consistently solved the $\Delta(T, B)$ of electron-doped WS$_2$ monolayer (Fig. S2a), which was experimentally reported to possess the highest $B_c^{Ising}/B_p$ ratios (~20) [6]. Specifically, the WFs were constructed using the $d$ orbital of W and $p$ orbital of S atom as the initial guess for the unitary transformations. We set $\langle\omega\rangle_{\log}$ equal to the Debye temperature (~210K) of WS$_2$ monolayer [10]. The pairing strength $g$ is set to 0.415, to reproduce the experimentally reported $T_c$ (~1.54 K) of WS$_2$ monolayer [6]. Then, the superconducting gap was solved self-consistently using a dense 400×400×1 **k**-points sampling in first Brillouin zone (Fig. S2b). These data enable us to extract the relation of $B_c/B_p$ versus $T/T_c$ (Fig. S2c), which can be reasonably fitted by using the microscopic model with the same $\beta_{SO}^*$=19.5 meV as that used in the experimental report (blue solid line) [6]. This indicates our self-consistently solutions for WS$_2$ agree with the experiment, and thus demonstrates the reliability of our predictions on the critical magnetic field of the MA$_2$Z$_4$ monolayers.

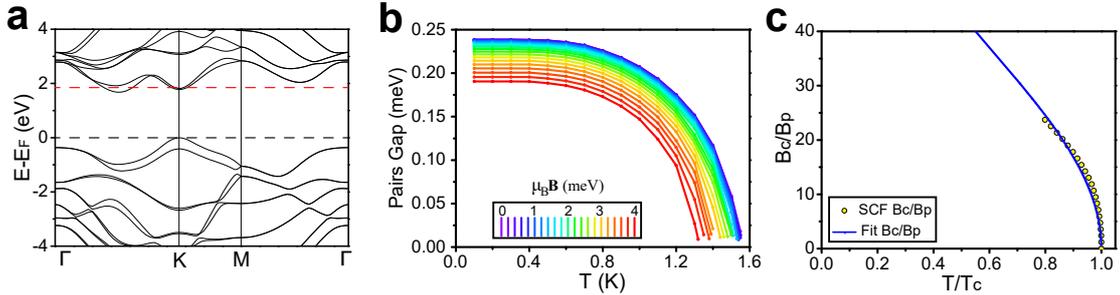

**Figure S2. Superconductivity of electron-doped WS$_2$ monolayer under magnetic field. a**, The electronic band structure of WS$_2$ monolayer. The black horizontal dashed line represents the Fermi level, and the red line is the energy position where the IBCS pairs are assumed to condensate. **b**, The temperature-dependent superconducting gap Δ of WS$_2$ monolayer under different in-plane magnetic fields, calculated self-consistently by using our newly developed method. **c**, The comparison between the relation of $B_c/B_p$ versus $T/T_c$ calculated self-consistently (yellow filled circles) and fitted by using the microscopic model with $\beta_{SO}^*$=19.5 meV (blue solid line).